\documentclass[10pt]{article}
\usepackage[utf8]{inputenc}
\usepackage[multiple]{footmisc}
\usepackage{caption}
\usepackage{graphicx}
\usepackage{subcaption}
\usepackage{amsmath}
\usepackage{amssymb}
\usepackage{mhchem}
\usepackage[margin=1in]{geometry}
\usepackage{color}
\usepackage[colorlinks,bookmarks,urlcolor=blue,citecolor=blue,linkcolor=blue]{hyperref}
\usepackage{authblk}
\usepackage{siunitx}

\title{Limited processivity of single motors improves overall transport flux of self-assembled motor-cargo complexes}
\date{}

\author[1,2]{Keshav B. Patel}
\author[2]{Shengtan Mao}
\author[1,2,3]{M. Gregory Forest}
\author[1,4,5]{Samuel K. Lai}
\author[6]{Jay M. Newby}
\affil[1]{UNC/NCSU Joint Department of Biomedical Engineering, University of North Carolina at Chapel Hill}
\affil[2]{Department of Mathematics, University of North Carolina at Chapel Hill}
\affil[3]{Department of Applied Physical Sciences, University of North Carolina at Chapel Hill}
\affil[4]{Division of Pharmacoengineering and Molecular Pharmaceutics, Eshelman School of Pharmacy, University of North Carolina at Chapel Hill}
\affil[5]{Department of Microbiology and Immunology, University of North Carolina at Chapel Hill, Chapel Hill}
\affil[6]{Department of Mathematical and Statistical Sciences, University of Alberta}

\begin{document}

\maketitle

\begin{abstract}
    Single kinesin molecular motors can processively move along a microtubule (MT) a few micrometers on average before dissociating.
    However, cellular length scales over which transport occurs are several hundred microns and more.
    Why seemingly unreliable motors are used to transport cellular cargo remains poorly understood.
    We propose a new theory for how low processivity, the average length of a single bout of directed motion, can enhance cellular transport when motors and cargoes must first diffusively self assemble into complexes.
    We employ stochastic modeling to determine the effect of processivity on overall cargo transport flux.
    We show that, under a wide range of physiologically relevant conditions, possessing ``infinite" processivity does not maximize flux along MTs.
    Rather, we find that low processivity i.e., weak binding of motors to MTs, is optimal. 
    These results shed light on the relationship between processivity and transport efficiency and offer a new theory for the physiological benefits of low motor processivity.
\end{abstract}

\section{Introduction}

Molecular motors function though a repeating sequence of reactions called a chemo-mechanical cycle.
Due to their size and scale of force generation, thermal fluctuations significantly influence single motor dynamics, making their activity and motion stochastic.
A motor's processivity is defined as the average distance that directed motion along a cytoskeletal filament is consistently engaged through the motor's chemo-mechanical cycle.
Disengagement from the filament is said to occur if the motor diffuses away from the filament into the cytoplasm and must randomly re-encounter a filament before directed motion can resume.
The processivity of single kinesin-1 (KIF5) motors, commonly seen in microtubule (MT) transport, has been measured to be $1-3\ \si{\micro\meter} $ \cite{}, while cargo transport distances, particularly in neurons, can be $10-100\ \si{\micro\meter}$ or more.
Another way of describing processivity is with the ``duty ratio" (denoted by $1-\alpha$ in this paper), which is defined as the average fraction of time an independent motor spends engaged to a filament.
Hence, the processivity is defined as the product of the duty ratio and the average motor velocity.

Processivity of the kinesin motor KIF5 was shown to be tunable through genetic modification of the neck linker; mutations that added positive charge increased processivity as much as 4-fold, while adding negative charge decreased processivity \cite{thorn2000}.
Even though mutations to the neck linker are capable of improving processivity, the neck coiled-coil regions are highly conserved. 
Taken together, these observations suggest that limited processivity is important for cell function \cite{thorn2000}.
If processivity can be increased as much as 4-fold with a simple single-nucleotide mutation, why does KIF5 have processivity in the range of a few microns when distances typical of active transport are often $10-100$ times longer? 
The purpose of this paper is to propose a new theory to answer the following question: is there a mechanistic purpose for limiting processivity of single motors for cellular transport?

Transport of self-assembled motor-cargo complexes has been shown to benefit from multiple-motor teams \cite{conway2012motor}.
A team of motors collectively has much higher processivity than individual motors.
While single motors may detach from MTs quickly, only one motor in the team needs to be engaged to maintain cargo transport.
Mathematical modeling shows that the mean time for the cargo to fully detach is an exponentially increasing function of the total number of motors in the team \cite{korn2009stochastic}.
However, these observations fail to explain why single motor processivity is limited when increasing it would dramatically improve performance of a multi-motor team.

Several theories for the role of limited processivity in active transport have been proposed. 
First, lower processivity may enhance the ability of motors to navigate obstructions and barriers \cite{thorn2000,conway2012motor}. 
Second, lower processivity of single motors may enhance their ability to effectively coordinate their chemo-mechanical cycles within a multi-motor team \cite{driver2011productive}.
Third, processivity may be tuned to achieve specific cargo and motor distributions across different cytoskeletal substrates \cite{rodionov1998functional}.

In this paper, we propose a new theory (which is not mutually exclusive to those previously proposed), namely, that limited processivity can improve the overall flux of cargoes when cargoes and motors must first diffusively self assemble into complexes before engaging in active transport.
Our conclusions are motivated by a recent theory for how antibodies trap virus in mucosal barriers by accumulating multiple antibodies on the viral surface that crosslink it to elements of the mucus polymer network \cite{newby2017blueprint}.
Our analysis shows that the improvement of cargo transport flux from lowering motor processivity is due to several competing factors.
Lower processivity increases the concentration of freely diffusing motors, which increases the rate of assembling multi-motor complexes from individual diffusive motors and cargoes, while simultaneously increasing the average number of motors per cargo complex.

The paper is organized as follows.
First, we introduce the stochastic model and discuss the underlying assumptions.
We then use a quasi-steady state (QSS) reduction of the model in two stages.
We present results using a Monte-Carlo solution to the full process, a numerical solutions of the partially reduced Chapman--Kolmogorov equation, and an analytical formula obtained from the full QSS reduction.

\section{Model}
Molecular motor transport can take place in a wide range of geometries e.g., spherical domains in transport of viral proteins to to the nucleus and branched domains found in neuronal dendrites \cite{newby2010Thesis,RevModPhys2013} and fungal syncytia. 
At the simplest level, motor transport occurs along linear filaments.
It is sufficient for our purposes to assume a one dimensional track of length $L$.
In our scenario, cargo begin at the negative end of the microtubule (MT) and travel down towards the positive end. 

The proposed theory depends on the following physical assumptions.
Cargoes and motors diffusively assemble into complexes in the cytoplasm before active transport of cargo can occur.
Single motors (either free or in the cargo complex) stochastically bind to and unbind from MTs.
For simplicity, we assume that when motors are bound to the MT, they move at a constant velocity ($\sim$1 \si{\micro\meter\per\second}). 
Multiple motors can bind to and unbind from a cargo, either in the cytoplasm or on MTs.
Cargoes diffuse in the cytoplasm if no motors are simultaneously bound to the cargo and MT.

We assume cargoes to be large (relative to motors), spherical packages that each have $N$ independent binding sites for motors. 
We define the following time dependent quantities. Let $n$ be the number of motors bound to a cargo and $s$ be the number of these $n$ motors that are also bound to a MT---in other words, the number of cargo-MT crosslinks.

Although the binding of motors and cargo is complex and requires multiple cofactors, we simplify the reaction to assume that the two are able to bind to each other directly. 
Binding and unbinding of motors and cargo occur with rates $k_{\rm on}$ and $k_{\rm off}$ respectively; the fraction of time motors spend unbound is given by $K=\frac{k_{\rm off}}{k_{\rm off}+k_{\rm on}}$. 
We assume that motors are sparsely distributed on the MT and thus it is unlikely for a motor bound to the MT to bind to a cargo that is already traveling along the MT (they can still encounter a free diffusive cargo), meaning the state change $(n,s)\to(n+1,s+1)$ is forbidden for $s>0$.
Moreover, a motor bound to the MT cannot diffusively explore the cytoplasm and therefore has a reduced binding rate (denoted as $k_{\rm on}^{\rm b}$) to diffusing cargo. 
By the Smoluchowski encounter rate, the reduction in the binding rate is $\frac{k^{\rm b}_{\rm on}}{k_{\rm on}} \propto \frac{D_{\rm cargo}}{D_{\rm cargo}+D_{\rm motor}}$. 

\begin{figure}[htbp]
    \centering
    \includegraphics[width=.4\textwidth]{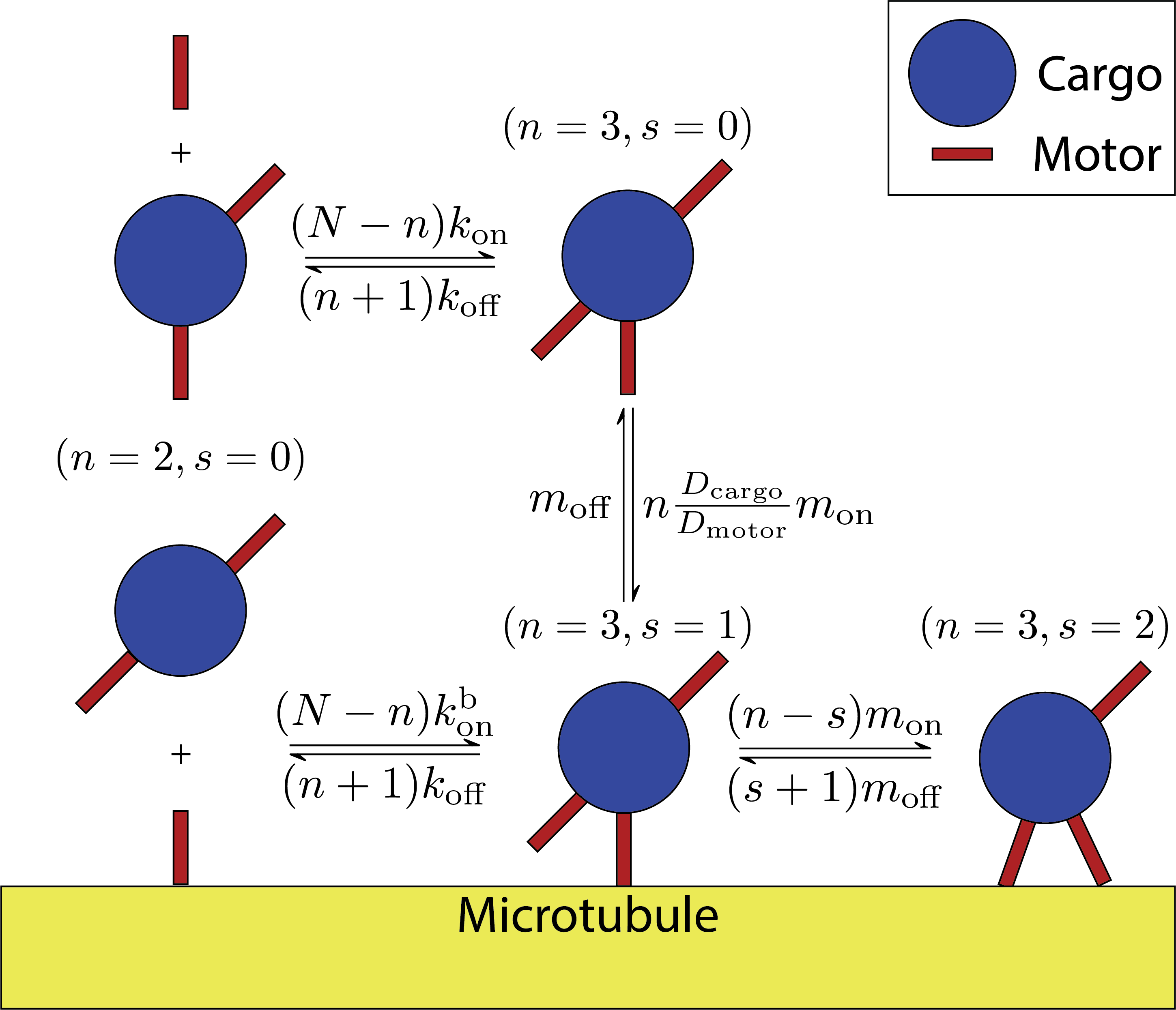}
    \caption{Binding diagram for reactions considered.}
    \label{fig:binding}
\end{figure}

Motors bind to and unbind from a MT with rates $m_{\rm on}$ and $m_{\rm off}$ respectively. 
Because the MT is essentially immobile, the binding rate between a motor of a free cargo and MT is reduced compared to the binding rate of free motors to a MT. In other words, we have that $\frac{m_{\rm on}^{\rm first}}{m_{\rm on}} \propto \frac{D_{\rm cargo}}{D_{\rm motor}}$.
When bound to a MT, the cargo moves at the same rate as a bound motor, and we assume the motor is unaffected by the cargo (i.e., we neglect mechanical interactions between the cargo and motors).

Assuming the background concentration of motors is at steady state, the fraction of motors free from the MT at any given time or position is given by $\alpha = \frac{m_{\rm off}}{m_{\rm off}  + m_{\rm on}}$. 

We use the following relations for binding constants: $k_{\rm on}=\alpha(D_{\rm cargo}+D_{\rm motor})\gamma$ and $k_{\rm on}^{\rm b}=(1-\alpha)D_{\rm cargo}\gamma$, where $\gamma$ is a scaling parameter related to the effective binding distance between motors and cargo. 
In this work, it is taken to be 3.57 to scale $k_{\rm on}$ to the appropriate value in Table \ref{tab:param} when $\alpha=1$ and small $D_{\rm cargo}$.

See Table \ref{tab:param} for a list of experimental values and ranges and Fig.~\ref{fig:binding} for a summary of the reactions considered.
{\renewcommand{\arraystretch}{1.5}
\begin{table}[h]
    \centering
    \begin{tabular}{|c|c|c|c|} \hline
        Parameter & Interpretation & Value(s) & References \\ \hline
        $N$ & number of motor binding sites on cargo & $5-10$ & \cite{Gross2007TwoMotors} \\ \hline
        $v_{\rm trans}$ & velocity of motor along MT & $.1-1\ \si{\micro\meter\per\second}$ & \cite{Bannai2004Vel,Klumpp2005Coop} \\ \hline
        $D_{\rm cargo}$ & diffusivity of transported particle & $10^{-4}-1\ \si{\micro\meter\squared\per\second}$ & NA \\ \hline
        $D_{\rm motor}$ & diffusivity of molecular motor protein & $1.4\ \si{\micro\meter\squared\per\second}$ & \cite{Grover2016Diff} \\ \hline
        $k_{\rm on}$ & binding rate between motors and cargo & $5\ \si{\Hz}$ & \cite{leduc2004cooperative} \\ \hline
        $k_{\rm off}$ & unbinding rate between motors and cargo & $1\ \si{\Hz}$ & \cite{Hancock1999Process} \\ \hline
    \end{tabular}
    \caption{Values and ranges of constants used in the model}
    \label{tab:param}
\end{table}}

The stochastic process is fully defined by the following Chapman--Kolmogorov equation:
\begin{equation}
\label{eq:ck_full}
    \frac{\partial}{\partial t}p(n,s,x,t) = \delta_{s,0}D_{\rm cargo}\frac{\partial^2 p}{\partial x^2} - (1-\delta_{s,0})v_{\rm trans}\frac{\partial p}{\partial x } + [M_s+V_{n,s}]p,
\end{equation}
where $p()$ is a probability density function describing cargo position at spatial position $x$, time $t$, where $n$ and $s$ are the number of bound motors and active crosslinks, respectively.
$M$ and $V$ are rate matrices describing motor/MT binding and motor/cargo binding respectively (see \eqref{M_s} and \eqref{V_n,s} in the Appendix for definitions).
The initial condition is given by
\begin{equation}
    p(n, s, x, 0) = \delta_{n, n_0}\delta_{s, s_0}\delta(x).
\end{equation}
We will consider two sets of boundary conditions: periodic
\begin{equation}
    \label{eq:bc1}
    p(n, s, 0, t) = p(n, s, L, t),\quad 
    \frac{\partial}{\partial x}p(n, 0, 0, t) = \frac{\partial}{\partial x}p(n, 0, L, t),
\end{equation}
and right-end absorbing
\begin{equation}
     \label{eq:bc2}
    \frac{\partial}{\partial x}p(n, 0, 0, t) = 0,\quad  p(n, s, L, t) = 0.
\end{equation}

\subsection{Quasi-Steady State Approximation}
In this section we enumerate the different characteristic timescales present in the system and their effect on overall transport efficiency.
Understanding the effect of separation between various timescales informs our physical understanding of the system and provides accurate approximations. 

We define the following timescale separation:
\begin{equation}
\tau_{\alpha} \ll \tau_K \ll \tau_{v},
\end{equation}
where $\tau_{\alpha} = \frac{1}{m_{\rm on} + m_{\rm off}}$ is the timescale of binding between motors and MT, $\tau_K = \frac{1}{k_{\rm on} + k_{\rm off}}$ is the timescale of binding between motors and cargo, and $\tau_v = \frac{L}{v_{\rm trans}}$ is the timescale of cargo transport. An additional implicit assumption for directed active transport is that cargo diffusion over transport length scales takes much longer than active transport. In other words, we assume that $\tau_{v} \ll \tau_{D}$, where $\tau_D = \frac{L^2}{2D_{\rm cargo}}$ is the timescale of cargo diffusion.

To begin, we can rewrite the full density function $p()$ as the product of conditional probabilities and the marginal density, $u(x,t)$:
\begin{equation}
    p(n,s,x,t) = \rho(s,t|n,x) \rho(n,t|x) u(x,t).
\end{equation}
 Because binding and unbinding are independent of spatial position, the conditional probabilities are independent of $x$. We assume further that $s$ and $n$ reach equilibrium rapidly so that we can assume that the marginal distributions are approximately at quasi-steady state. Hence, we have that $\rho(s,t|n,x) \sim \rho(s|n)$ and $\rho(n,t|x) \sim \rho(n)$. These are the quasi-steady state probability distributions, and by definition we have that the net reaction rates are zero so that
\begin{equation}
    \label{eq:qss_def}
    \sum_{s=0}^N M_s\rho(s|n) = 0,\quad
    \sum_{n=0}^N\sum_{s=0}^N V_{n,s}\rho(s|n)\rho(n) = 0.
\end{equation}
It can be readily verified that the quasi-steady state distributions, satisfying \eqref{eq:qss_def}, are given by
\begin{equation}
\rho(s|n) = \begin{cases} \frac{D_{\rm motor}\alpha^n}{(D_{\rm motor}+D_{\rm cargo})\alpha^n+D_{\rm cargo}}, & s = 0 \\[10pt]
\frac{D_{\rm cargo}\binom{n}{s}(1-\alpha)^s\alpha^{n-s}}{(D_{\rm motor}+D_{\rm cargo})\alpha^n+D_{\rm cargo}}, & 0 < s \leq n \end{cases}
\label{PsGivenn}
\end{equation}
and
\begin{equation} 
\rho(n) = \frac{\mathcal{N}}{k_{\rm off}^n}\binom{N}{n}\prod_{j=0}^{n-1}(\rho(0|j)k_{\rm on}+k^{\rm b}_{\rm on}).
\label{Pn}
\end{equation}

Applying the QSS assumption, we obtain the following reduced description of motor transport. Substituting $p(n,s,x,t) \sim \rho(n|s)\rho(n)u(x,t) + O(\epsilon)$ into \eqref{eq:ck_full} and summing over $s$ and $n$ yields
\begin{equation}
     \frac{\partial}{\partial t}u(x,t)=\overline{D}\frac{\partial^{2} u}{\partial x^2} - \overline{v}\frac{\partial u}{\partial x},
    \label{ck_qss}
\end{equation}
where
\begin{equation}
    \overline{D} = D_{\rm cargo}\sum_{n=0}^N \rho(0|n)\rho(n), \quad 
    \overline{v} = v_{\rm trans}\sum_{n=0}^N\sum_{s=1}^n \rho(s|n)\rho(n).
\end{equation}
The reduced parameters, $\overline{D}$ and $\overline{v}$, are the steady-state diffusivity and transport velocity respectively of a population of cargo, and $\mathcal{N}$ is a normalization factor. Eqn.~\eqref{ck_qss} is then solved over the entire real line using the Fourier Transform and truncated to $(-\infty, L]$ by the Method of Images to arrive at a closed-form approximation to the solution:
\begin{equation}
    u(x,t)=\frac{1}{\sqrt{4\pi \overline{D}t}}\bigg(e^{-\frac{(x-\overline{v}t)^2}{4\overline{D}t}}-e^{\frac{\overline{v}L}{\overline{D}}-\frac{(x-\overline{v}t-2L)^2}{4\overline{D}t}
}\bigg).
\label{analytic}
\end{equation}
We then truncate the domain to $[0,L]$ so that we can compare with our other scales. See Appendix for a discussion of the error in this assumption.

It is informative to examine the system at different levels of approximation. 
An intermediate QSS approximation that assumes only that $\tau_{\alpha}\ll \tau_{K}$ is obtained by condensing the system of two binding reactions to one effective binding reaction between motors and cargo. 
Denote the probability that a cargo is freely diffusing ($s=0$) given that it has $n$ motors attached is given by $\beta(n) = \rho(0, n)$, where $\rho(\cdot, \cdot)$ is defined in \eqref{eq:qss_def}.
The effective binding and unbinding of a cargo and motor is summarized in Eq. \eqref{QSSrxn}:
\begin{equation}
    M+M^nC\ce{<=>[$(N - n)(k_{\rm on}\beta(n) + k^{\rm b}_{\rm on})$][$(n + 1) k_{\rm off}$]}M^{n+1}C
    \label{QSSrxn}
\end{equation}
Where $M$ denotes a motor and $C$ denotes a cargo. 
We use this binding reaction along with the diffusive and transport terms to define a one-dimensional partial differential equation given by Eq. \eqref{PDE} (See Appendix).
Note that the two approximations (\eqref{PDE} and \eqref{ck_qss}) are consistent; if we apply an additional timescale separation $\tau_{K} \ll \tau_{V}$ then \eqref{PDE} converges to \eqref{ck_qss}.

\subsection{Transport Flux and Mean Transit Times}

We are interested in exploring parameter conditions that allow for effective transport. 
To quantify this we examine the probability flux under our two boundary conditions \eqref{eq:bc1} and \eqref{eq:bc2}. 
For an absorbing boundary at the end of the microtubule, we have the time-dependent flux given by
\begin{equation}
    J(t) = \left. -D_{\rm cargo}\sum_{n=0}^{N}\frac{\partial p}{\partial x}\right|_{s=0,x=L}
    + v_{\rm trans} \sum_{n=1}^{N}\sum_{s=1}^{n}p(n, s, L, t).
\end{equation}
The survival probability, defined as the probability that the cargo has not exited the system at $x=L$ before time $t$, is given by
\begin{equation}
    S(t) = \int_{0}^{t} J(t')dt'.
\end{equation}
The median transport time $T_{\rm median}$ is given implicitly by $S(T_{\rm median}) = 0.5$.
For periodic boundary conditions, we examine the steady-state probability flux given by
\begin{equation}
    J_{\infty} = \lim_{t\to\infty}  v_{\rm trans} \sum_{n=1}^{N}\sum_{s=1}^{n}p(n, s, L, t).
\end{equation}
Based upon our QSS approximation \eqref{ck_qss}, the steady state flux is approximated by $J_{\infty} \sim \frac{\bar{v}}{L}$.

\section{Results}

\paragraph{Cooperation among low processivity motors optimizes cargo flux}

\begin{figure}[htbp]
    \centering
    \includegraphics[width=\textwidth]{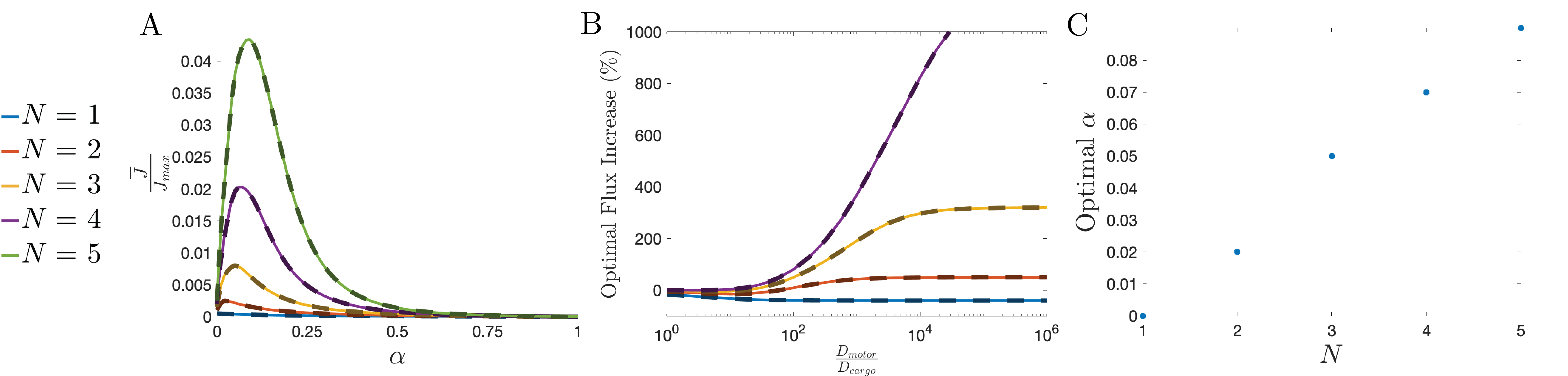}
    \caption{{\bf The steady state behavior as a function of $\alpha$, the fraction of freely diffusing motors.} (A) Effective transport flux, calculated using the velocity of the cargo population once net binding rate between cargo and motors reached 0 ($D_{\rm cargo}=0.14\cdot 10^{-3}\ \si{\micro\meter\squared\per\second}$). 
    (B) Maximum \% increase in steady-state effective flux when motors are allowed to dissociate from MTs during transport. 
    (C) $\alpha$ values at the peaks in (A) as a function of $N$. Solid lines denote the QSS solution and dashed lines denote the computational solution.
    Other parameters used for both cases are: $L=10\ \si{\micro\meter}$, $v_{\rm trans}=1\ \si{\micro\meter\per\second}$, $k_{\rm on}=5\ \si{\Hz}$, $k_{\rm off}=1\ \si{\Hz}$ and $D_{\rm motor}=1.4\ \si{\micro\meter\squared\per\second}$}
    \label{fig:numSitesFluxIncrease}
\end{figure}

To obtain a steady state cargo flux, we impose periodic boundary conditions and begin with cargo evenly distributed throughout the system. 
We allow the system to approach a steady-state distribution by stopping the simulation when the change between each time step is sufficiently small.
We then calculate the average flux and compare it to the theoretical maximum.
Figure \ref{fig:numSitesFluxIncrease}(A), which showed a significant change in transport flux with respect to $\alpha$. 

When $N>1$, we see an optimum is present around $\alpha=0.1$, where on average $\sim 10\%$ of motors are freely diffusing in the cytosol at any given time.
Figure \ref{fig:numSitesFluxIncrease}(C) plots the location of this optimum.
Increasing $N$, the total number of binding sites for motors, causes an increase in the effective transport flux for all values of $\alpha$.
Surprisingly, limited processivity leads to larger steady-state transport fluxes than using perfectly processive motors that do not dissociate from the MT ($\alpha=0$), provided that cargoes possess multiple motor binding sites (i.e., $N>1$).
To gauge the robustness of the optimal flux, we calculate the percentage change in flux between $\alpha=0$ and $\alpha=0.1$ for different values of $N$ (see Fig.~\ref{fig:numSitesFluxIncrease}(B)).
Even for a relatively modest number of motors ($N=3$), optimizing the processivity yields a flux increase in excess of 300\% over perfectly processive motors when the diffusivity of the free motor protein is high relative to the diffusivity of the free cargo. 

\paragraph{A separation of motor binding timescales reduces cargo passage times}
\begin{figure}[tbp]
 \begin{minipage}{\textwidth}
    \centering
    \includegraphics[width=.49\textwidth]{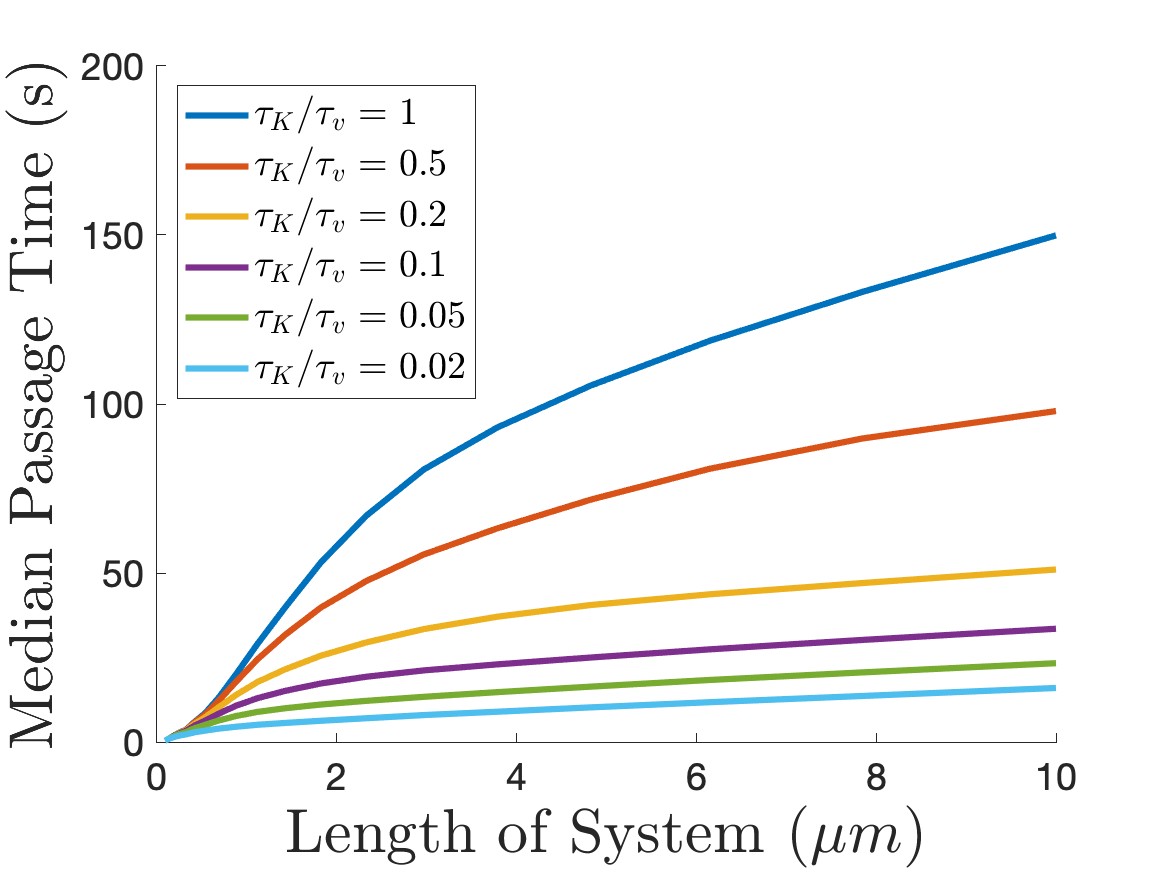}
     \includegraphics[width=.49\textwidth]{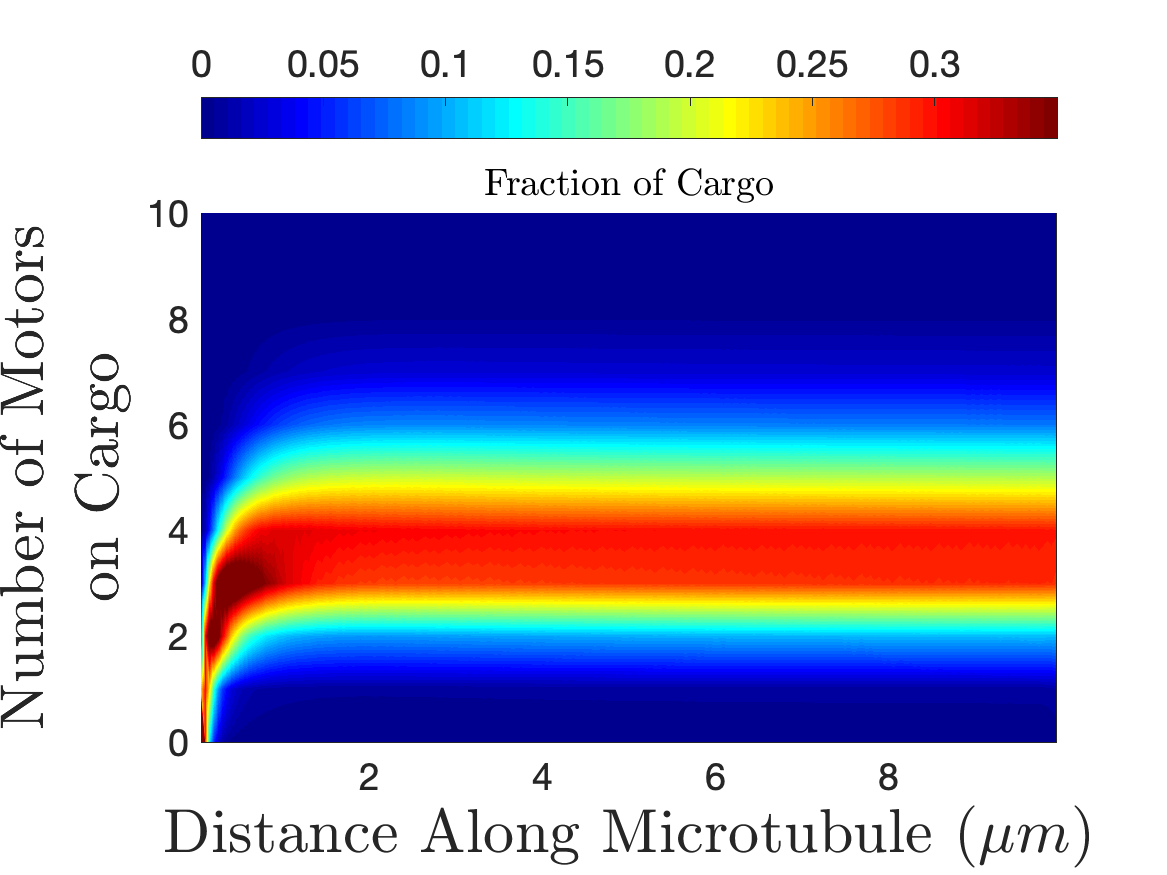}
      \includegraphics[width=.49\textwidth]{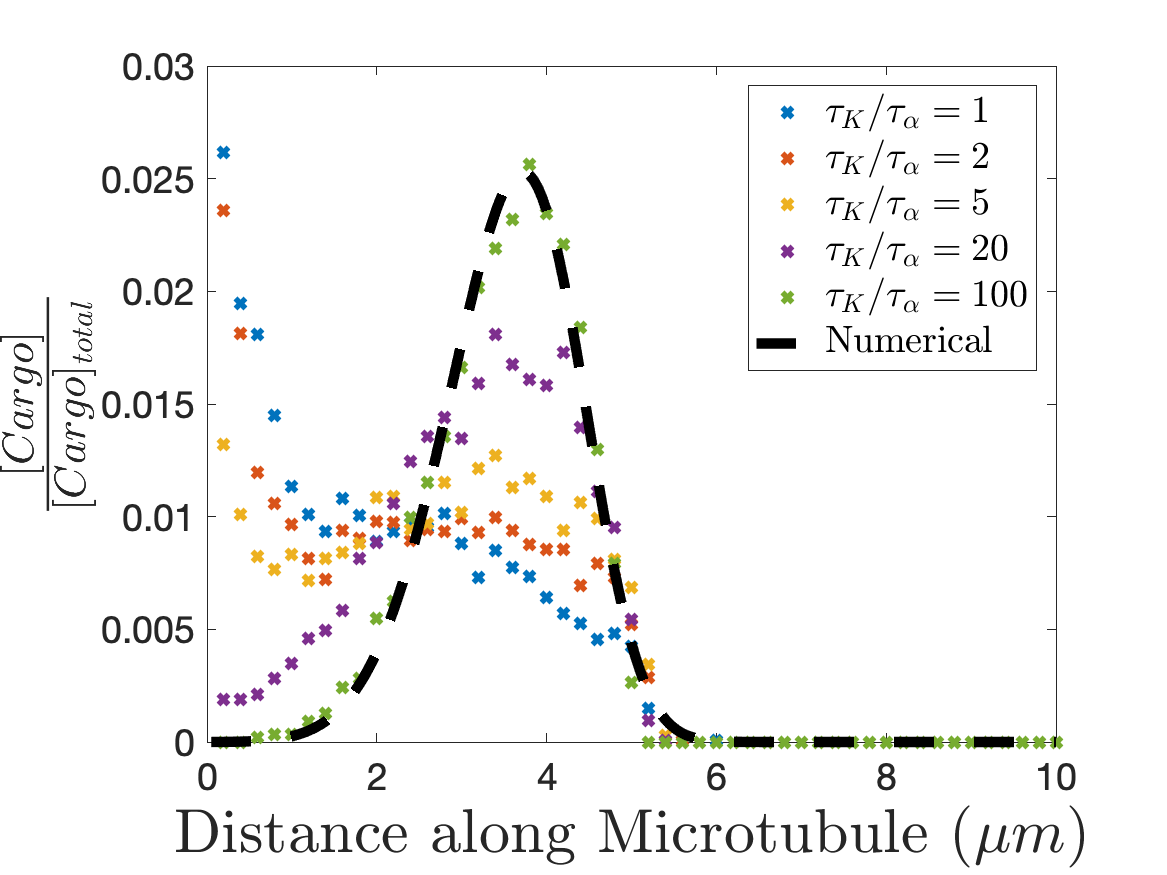}
     \includegraphics[width=.49\textwidth]{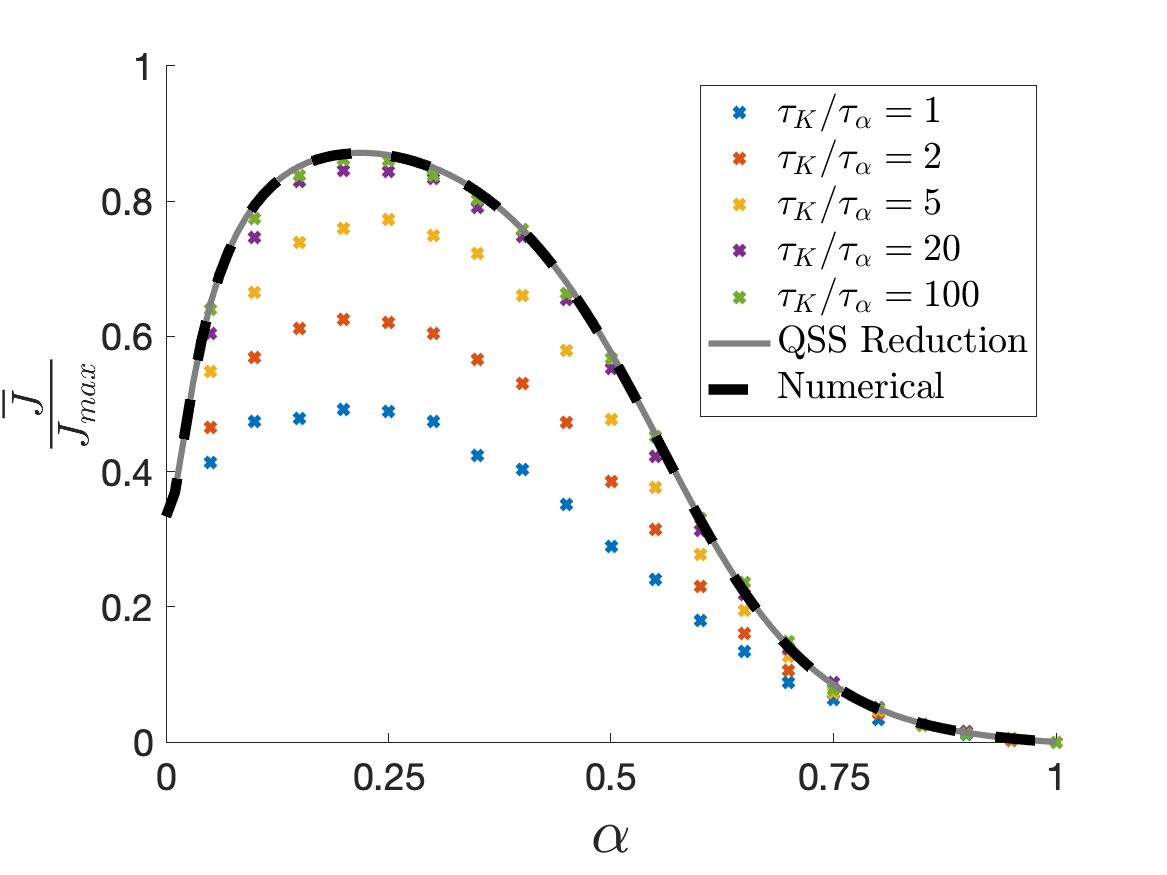}
\end{minipage}
    \caption{{\bf The effect of timescale separation on transport efficiency.} (A) The median time to reach the end of the system. (B) The distribution of motors on the cargo (at the median time) as a function of distance along the MT.
    (C, D) Convergence between the Monte Carlo simulation and computational PDE solution as quasi-steady state assumption is taken. Constants used are: $N=10$, $D_{\rm cargo}=0.014\ \si{\micro\meter\squared\per\second}$, $D_{\rm motor}=1.4\ \si{\micro\meter\squared\per\second}$, $v_{\rm trans}=1\ \si{\micro\meter\per\second}$, $\alpha=0.1$, $k_{\rm on}=5\ \si{\Hz}$, $k_{\rm off}=1\ \si{\Hz}$.
    }
    \label{fig:timeToClear}
\end{figure}

Cargo are assumed to not have any motors bound to them when they begin at the negative end of the MT. 
As such, we expect our binding parameters to play a significant role in how quickly cargo reach a steady-state equilibrium with the rest of the system.
Here, shorter median passage times correlate with an increase in the system's efficiency in moving cargo.
In the following studies, we impose a Robin boundary condition with all cargo beginning at position $x=0$ and monitor cargo distributions up to the absorbing boundary.

The time needed for cargo to equilibrate with motors also affects transit times. 
We use the aforementioned setup to calculate the amount of time needed for the probability of passage to the end of the domain to reach $0.5$ (i.e., the median passage time; Fig.~\ref{fig:timeToClear}(A).
At very small length scales (i.e. $<1\ \si{\micro\meter}$) where diffusion dominates, binding timescales smaller than that of transport results in shorter travel times. 
However, this is quickly overcome, such that after approximately $2\ \si{\micro\meter}$, faster binding result in shorter median passage times.
In Fig.~\ref{fig:timeToClear}(B), we record the distribution of motors on the cargo as a function of position $x$ at the median time.
We find that cargo rapidly bind to motors as they move down the system, such that most cargo have more than one motor and the distribution reaches quasi-steady state within a few microns. 

Fig.~\ref{fig:timeToClear}(C) plots the convergence to the QSS assumption. As $\tau_{\alpha}$ decreases (i.e. more rapid binding and unbinding as reflected by an increase in $m_{\rm on}$ and $m_{\rm off}$) we see that our solution rapidly converges to the numerical solution, the regime that maximizes cargo flux along MT. 
In Fig.~\ref{fig:timeToClear}(D), we see that decreasing $\tau_{\alpha}$ creates the optimum in transport flux at $\alpha=0.2$. 
From these figures we conclude that short-lived binding between motors and MT, as seen physiologically, is in fact a necessary condition to create an optimum transport flux.

\paragraph{Freely diffusing motors bind cargo more rapidly than motors engaged on a MT}
\begin{figure}[htbp]
    \centering
    \includegraphics[width=.9\textwidth]{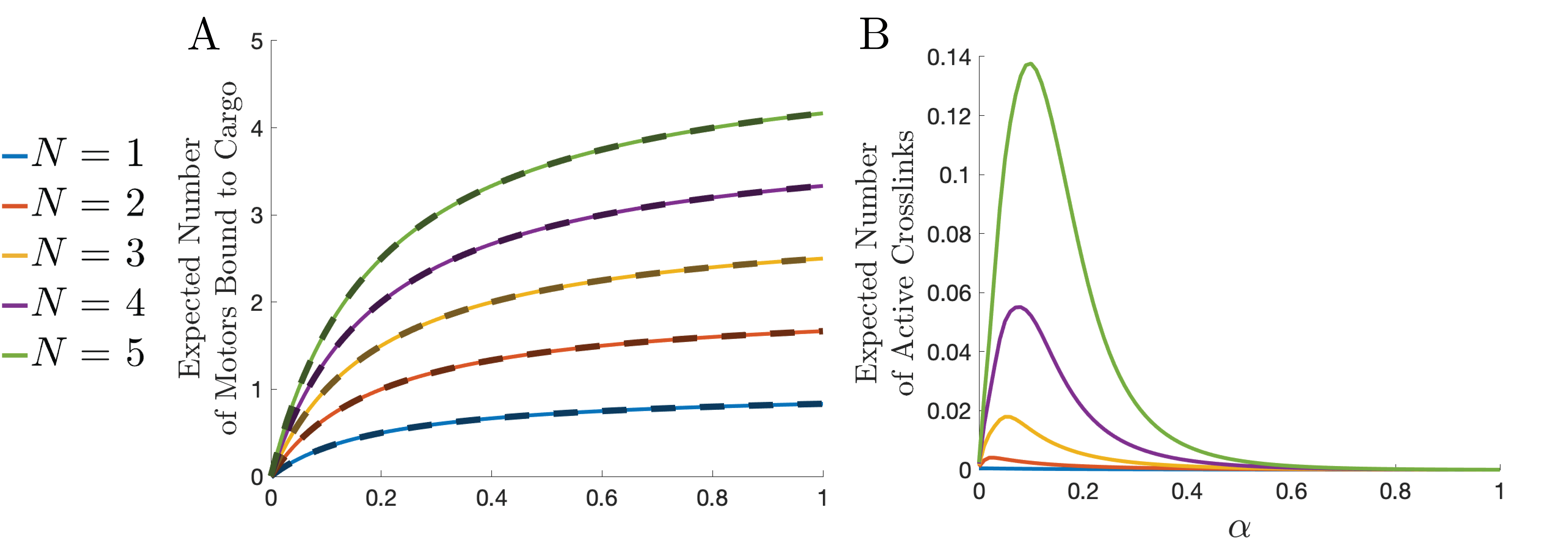}
    \caption{{\bf The average number of motors and average number of engaged motors at steady state.} Solid lines denote the the QSS solution and dashed lines denote the computational solution. (A) Expected number of motors on cargo. (B) Expected number of motors bound to both cargo and MT. Same parameter values as used in Figure \ref{fig:numSitesFluxIncrease}.
    }
    \label{fig:transportPlotsNumsites}
\end{figure}
To understand the relationship between processivity and transport, we look at the average number of motors on the cargo. 
In the same model setup used to create Fig.~\ref{fig:numSitesFluxIncrease}, we use the distribution of the number of motors bound to cargo at steady-state to calculate averages. 
We also use the analytic equation for the conditional probability given by \eqref{PsGivenn} to calculate the average number of crosslinks.

The optimum seen in Fig.~\ref{fig:transportPlotsNumsites}(B) for $N>1$ is a result of two competing factors.
The average distance a motor travels before detaching increases as $\alpha\to0$.
On the other hand, maximizing the total number of motors bound to the cargo (see Fig.~\ref{fig:transportPlotsNumsites}(A)) requires low processivity ($\alpha\to1$), because the encounter rate is higher between freely-diffusing motors and much less mobile cargo.

Based on the preceding discussion, we conclude that cooperation among molecular motors robustly enhances transport flux, which can be further optimized by tuning processivity.
However, motors must be allowed sufficient time to bind and accumulate on the cargo so that once the complex encounters a MT, many motors are able to cooperate during transport.

\section{Discussion}
We have presented a new theory to compliment the expanding literature regarding motor processivity. 
The theory's major benefits are two-fold.
First, it is based on simple, well documented motor binding kinetics, providing an easily explained mechanism.
Second, the results are based on a broadly defined model that connects to the more general theory that fast, weak-affinity interactions of crosslinking molecules provide optimal steady-state binding characteristics (for example, between antibodies and mucin polymers) \cite{newby2017blueprint, Wang2014Mucus}.

Before a cargo accumulates motors and encounters a MT, the dominant mode of travel is diffusion. 
Diffusive motion is not efficient for transport over large distances, but it is efficient for self assembly of cargo-motor complexes.
Once a sufficient number of motors bind to the cargo and a MT is encountered, directed motion becomes dominant and diffusive motion has little effect on the remaining travel time to the end of the MT.

Allowing multiple motors on a cargo exponentially increases the probability of maintaining at least one motor engaged to the MT at all times. 
This is true for any fixed $0<\alpha<1$ (the fraction of freely diffusing motors).
As motor processivity is increased ($\alpha \to 0$), there are many motors on MTs but very few on freely diffusing cargo. 
On the other hand, as $\alpha \to 1$, there are potentially many motors bound to a cargo, but few motors engaged to a MT.
These competing factors set the stage for an optimal transport flux for an intermediate value of $0 < \alpha< 1$ (we observe $\alpha \approx 0.1$ is optimal for reasonable parameter choices). 

We have also shown that cargo with smaller diffusivities (larger hydrodynamic radius) compared to individual motors benefit more from low-processivity motors than smaller cargoes with higher diffusivity.
Because bimolecular binding rates are diffusion-dependent, smaller diffusivities lead to lower binding rates. 
The binding rate ($k_{\rm on}^{\rm b}$) between engaged motors (on a MT) and freely diffusing cargo is proportional to $D_{\rm cargo}$.
The binding rate ($k_{\rm on}$) between freely diffusing motors and cargo is proportional to $D_{\rm cargo}+D_{\rm motor}$, which is comparatively unaffected by cargo size and diffusivity when $D_{\rm cargo} \ll D_{\rm motor}$. 
As a result, motor-cargo binding at $\alpha=0$ (where all motors are engaged to a MT) becomes an unlikely event for large cargoes with small diffusivity.
While not explored here, additional factors such as confinement and subdiffusion in a visco-elastic liquid can only compound this effect.

Agreement between the three timescales of our model was instrumental for validating numerical codes and to highlight the convergence of our approximations.
In particular, convergence between full and reduced models as the motor-MT binding timescales decreases shows that a single diffusion equation can describe the system under exactly the conditions that make the optimal transport phenomenon possible (i.e., the separation of timescales). 
Finally, we considered the simplest geometry of active transport, namely, a 1D MT domain of length $L$.
Future work will be necessary to elucidate the additional effects of more complicated geometries on the optimal transport flux phenomena introduced here.

\section{Appendix}

\paragraph{Numerical Computation of PDE}

Under the quasi steady state assumption, the system can be equivalently modeled by examining the population of cargo using the following 1-D partial differential equation:

\begin{equation}
    u_t(x,t) = \mathbf{D}u_{xx} - \mathbf{V}u_x + \mathbf{A}u
    \label{PDE}
\end{equation}
where the vector $u$ is an $N+1$ vector of cargo concentrations and the $i^\text{th}$ component corresponds to cargo with $i$ motors bound. $\mathbf{D}$ and $\mathbf{V}$ are diagonal matrices with $\mathbf{D}_{ii} = \beta(i)D_{\rm cargo}$ and $\mathbf{V}_{ii} = (1 - \beta(i))v_{\rm trans}$. 
$\mathbf{A}$ has values on the off-diagonals corresponding to the reactions in Equation \eqref{rxn1}.
In this work, \eqref{PDE} is solved in MATLAB and data is stored as concentrations and fluxes at each position and time.
The Forward-Time Centered-Space method is used to solve the diffusion equation (Forward-space for the advection term).
Boundary conditions are usually considered to be Robin and initial conditions are set as a cargo concentration with $n=0$ at $x=0$ only.
The stop criterion is determined by the amount of cargo still in the system.

\begin{figure}[htbp]
    \centering
    \includegraphics[width=.5\textwidth]{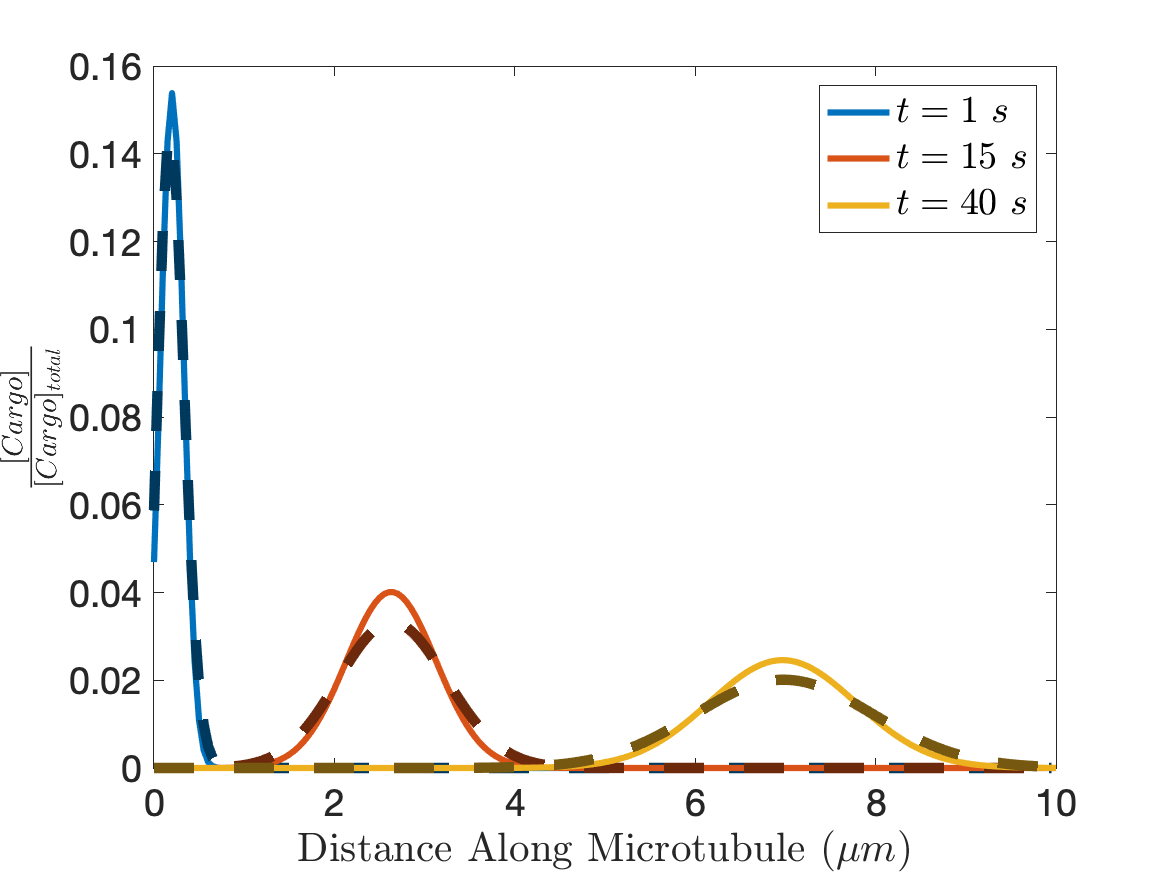}
    \caption{Comparison of Steady-state computational scheme and analytic scheme defined by Equation \eqref{analytic}. Constants used: $N=10$, $L=10\ \si{\micro\meter}$, $v_{\rm trans}=1\ \si{\micro\meter\per\second}$, $k_{\rm on}=.2\ \si{\Hz}$, $k_{\rm off}=1\ \si{\Hz}$, $D_{\rm cargo}=.01\ \si{\micro\meter\squared\per\second}$, and $D_{\rm motor}=10\ \si{\micro\meter\squared\per\second}$}
    \label{fig:FTCSvsCK}
\end{figure}

However, for steady-state analyses we choose a periodic boundary condition and a uniformly distributed initial condition of cargo with $n=0$. 
The stop criterion in this setup is determined by the magnitude of the change in the distribution of cargo concentrations with respect to $n$.
See Fig.~\ref{fig:FTCSvsCK} for a comparison between the computational and analytic methods.
The solution is best for large $\overline{v}$ because the effect of the absorbing boundary condition at $x=0$ is minimized. 
Therefore, for a given diffusivity and transport velocity, using the optimal $\alpha$ will create the best congruence between the two schemes.
At worst (i.e. $\overline{v}=0$), the analytic solution is half the value of the computational model for all $x$.
Because the focus on this work was primarily on the agreement between the analytic and computational $\overline{v}$, performing a more exact calculation was not necessary.
It is possible to perform additional iterations of the Method of Images to improve the analytic solution. 
This procedure can be done infinitely, creating a sequence of functions that converges to the same solution as the computational scheme (up to machine precision).

\paragraph{Monte Carlo Simulation}

The Monte Carlo Simulation no longer assumes the binding and unbinding rates between microtubule and motors are infinite. 
Let $m_{\rm on}$ and $m_{\rm off}$ denote those finite rates.
Let $k_{\rm on}$, $k_{\rm on}^{\rm b}$, and $k_{\rm off}$ be as before.

For a single cargo, let $N$ be its total number of binding sites.
Let $n$ denote the number of motors bound to the cargo, and let $s$ denote the number of those bound motors that are also cross-linked. 
Let the state be defined as the ordered pair $(s,n)$. 
The state changes according to the following rates: 
\begin{equation}
    (s,n)\ce{<=>[$(N - n)k_{\rm on}$][$(n-s+ 1) k_{\rm off}$]}(s,n+1)
    \label{rxn1}
\end{equation}
\begin{equation}
    (s,n)\ce{<=>[$\delta_{s,0}(N - n)k_{\rm on}^{\rm b}$][$(s+1) k_{\rm off}$]}(s+1,n+1)
    \label{rxn2}
\end{equation}
\begin{equation}
    (s,n)\ce{<=>[$g(s)(n - s)m_{\rm on}$][$(s+1) m_{\rm off}$]}(s+1,n)
    \label{rxn3}
\end{equation}
\begin{equation}
g(s)=
\begin{cases}
    \frac{D_{\rm cargo}}{D_{\rm motor}},& \text{if } s= 0\\
    1,              & \text{else}
\end{cases}
\end{equation}

With these state changes, the reaction rate matrices $M_s$ and $V_{n,s}$ are the following $N+1\times N+1$ matrices:

\begin{equation}
    M_s=\begin{bmatrix}
        0                                  & m_{\rm off}     & 0           & \dots & & 0 \\
        n\frac{D_{\rm cargo}}{D_{\rm motor}}m_{\rm on} & 0           & 2m_{off}    & \dots & & 0 \\
        0                                  & (n-1)m_{\rm on} & 0           & \dots & & 0 \\
        0                                  & 0           & (n-2)m_{\rm on} & \ddots & & 0 \\
        0                                  & 0           & 0           &        & & 0 \\
        \vdots                             & \vdots      & \vdots      &        & & nm_{\rm off} \\
        0                                  & 0           & 0           &        & & 0 \\
    \end{bmatrix}
    \label{M_s}
\end{equation}

\begin{equation}
    V_{n,s}=\begin{bmatrix}
        0                  & k_{\rm off}                            & 0           & \dots & & 0 \\
        N(k_{\rm on}+k_{\rm on}^{\rm b}) & 0                                  & 2k_{\rm off}    & \dots & & 0 \\
        0                  & (N-1)(k_{\rm on}+\delta_{s,0}k_{\rm on}^b) & 0           & \dots & & 0 \\
        0                  & 0                                  & (N-2)(k_{\rm on}+\delta_{s,0}k_{\rm on}^b) & \ddots & & 0 \\
        0                                  & 0           & 0           &        & & 0 \\
        \vdots                             & \vdots      & \vdots      &        & & nk_{\rm off} \\
        0                                  & 0           & 0           &        & & 0 \\
    \end{bmatrix}
    \label{V_n,s}
\end{equation}

The code first builds a square matrix where the rows and columns represent all possible states $(s,n)$.
The entry in $(s_1,n_1)$ row and $(s_2,n_2)$ column is the rate of $(s_2,n_2)$ going to $(s_1,n_1)$.
For the actual simulation, suppose our cargo is currently at position $x$, time $t$, and just changed to state $(s,n)$. 
Determine the time elapsed by 
\begin{equation} t_e=-\frac{1}{\lambda}log(U(0,1)) \end{equation}
where $\lambda$ is the rate to exit the current state.
If $s>0$, it means the cargo is cross-linked, so it is moving only by transport. 
The change in position
\begin{equation} \Delta x=vt_e \end{equation}
where $v$ is the transport velocity.
If $s=0$, it means the cargo is free-floating, therefore it is only moving by diffusion.
Using the equation 
\begin{equation} \Delta x=\sqrt{2D_{\rm cargo}t_e}N(0,1) \end{equation} 
we can determine the next position of the cargo.
We determine the next state by using the rates matrix and a uniform random variable. 
The next time is given by $t+t_e$, and the next position is given by $x+\Delta x$.

A few things to note: we start the cargo with $x=0, t=0, (s,n)=(0,0)$. 
The left boundary is reflective, and the right boundary is absorbing.
If the cargo goes into negative positions, it is adjusted to be at $x=0$.
If the cargo reaches the end of microtubule with length $l$, the simulation for this cargo is terminated.

By running this simulation for a large number of cargoes, we can extract data analogous to the data produced from the computational and analytic solutions. 
As Fig.~\ref{fig:timeToClear}(C) shows, if we let the values of $m_{\rm on}$ and $m_{\rm off}$ tend to infinity while keeping their ratio constant, the density distribution of the cargoes along the microtubule tends to the computational and analytic solutions at any given time.

\bibliography{refs}
\bibliographystyle{abbrv}
\end{document}